\documentclass[12pt]{article}
\usepackage{amssymb}
\usepackage{amsmath}
\usepackage{graphicx}
\def\kx{k_x}
\def\ky{k_y}
\def\kz{k_z}
\def\k0z{k_{0 z}}
\def\expo{e}
\def\kxs{k_x^2}
\def\omegas{\omega^2}
\def\kc{k_0}
\def\kcs{k_0^2}

\begin{document}
\title{
\vskip-2truecm{\hfill {\small CFNUL/01-03}} \\
\vskip0.001truecm{\hfill {\small DF/IST-2.2001}}
\vskip0.001truecm{\hfill {\small gr-qc/0104015}} \\
\vskip-3mm {\normalsize \hfill } \\
\vskip5mm Interaction between gravitational waves\\ and domain walls }
\date{March, 2001}
\author{Lu{\'{\i}}s Bento \\
\emph{{\normalsize Centro de F\'{\i}sica Nuclear da Universidade de Lisboa,}}
\\
\emph{\normalsize Avenida Prof. Gama Pinto 2, 1649-003 Lisboa, Portugal}\\
{\normalsize \ E-mail: lbento@fc.ul.pt}\\
\strut \and Jos\'{e} P. S. Lemos \\
\emph{{\normalsize CENTRA, Departamento de F\'{\i}sica, Instituto Superior
T\'{e}cnico,}}\\
\emph{\normalsize Avenida Rovisco Pais 1, 1049-001 Lisboa, Portugal }\\
{\normalsize \ E-mail: lemos@kelvin.ist.utl.pt}}

\maketitle

\begin{abstract}
We study the gravitational perturbations of thick domain walls.
The refraction index and spin properties of the
solutions interior to the wall are analyzed in detail.
It is shown that the gravitational waves suffer a refraction process 
by domain walls.
The  reflection and
transmission coefficients are derived in the thin wall limit.  
In relation to the spin content, it is shown that the
``$\times$''  helicity 2 gravitational
 wave mode maintains in the domain wall
the same polarization state
 as in vacuum.
On the contrary, the  ``$+$'' mode, of  pure helicity 2 in vacuum,
 is contaminated inside the
wall with a spin 0 state, as well as with spin 2, helicity 0 and 1 states.

\strut \newline
\end{abstract}

\newpage


\section{Introduction}  \label{introduction}

Topological defects occur in condensed matter systems and in field
theories of elementary particles. When the vacuum does not possess all
the symmetries of the theory, a symmetry breaking can occur, leading
to the formation of topological defects. This symmetry breaking is
usually described in terms of scalar fields.  In cosmology, this
breaking occurs spontaneously, when the early universe has cooled down
to some critical value of the temperature.  Depending on the type of
symmetry that is being broken, topological defects can be domain
walls, strings, monopoles and textures (see~\cite{vilenkinshellard} for
all details). 

In this work, we are interested in domain walls. Domain walls appear
through the breaking of a discrete symmetry. In the simplest
realization the vacuum has two states, in which the scalar field
$\varphi$ can take values $\varphi_+$ and $\varphi_-$, say. When, on
one side of two neighboring regions it takes the value $\varphi_+$ and
on the other side $\varphi_-$, then a domain wall occurs in the
separation layer, with the field $\varphi$ interpolating between these
two values. These are the $Z_2$ domain walls, which arise 
due to a discrete symmetry breaking of two possible states. 
The 3+1-dimensional (3+1D) domain walls, are in fact 
the kink (soliton) solutions in 1+1D when extended 
through the other two transversal directions (thus, n+1D domain wall 
solutions of this type  exist for n$\geq1$).

Since their introduction~\cite{kibble,vilenkin1}, domain walls 
have  been rejected and readmitted several times as possible 
objects inhabiting our universe. 
Initially they were ruled out because once formed, with a typical 
energy scale of $\gtrsim 1\,$MeV, they would dominate 
catastrophically the energy density of the universe. 
Nevertheless, they showed interesting gravitational effects, and 
within this perspective, were analyzed in detail using a thin wall 
approximation~\cite{vilenkin2,ipsersikivie}. 
Domain walls were resurrected in a new scenario where a possible 
late phase transition occurred~\cite{hillschrammfry}. The critical 
temperature here is low, and the wall can acquire a relative 
large thickness. These walls would produce very low density 
contrasts and lack any observational support. 
Domain walls appeared a third time when it was suggested 
\cite{vilenkin3} that Planck size topological defects
could trigger inflation. 
Recently, the study of domain walls reappeared in higher 
dimensions, after the suggestion that the universe is a 
three dimensional domain wall in a higher-dimensional universe
\cite{rubakov83,randallsundrum,akama00}.

Important aspects that have been discussed in the literature are the
gravitational interaction of the domain walls with other particles and
walls~\cite{vilenkinshellard,everett}, the dynamics of the 
walls~\cite{vilenkinshellard,bonjour1}, and the intrinsic structure of the
wall itself~\cite{vilenkinshellard,widrow89,goetz,ishibashi,bonjour2}.
The gravitational field of the wall has some interesting features.
For instance, both the interior and exterior gravitational fields are
not static. In addition, the gravitational field to external observers
is repulsive, an effect which can be seen differently according to the
coordinate system used~\cite{ipsersikivie}. 

Here we want to consider thick domain walls, of the kind studied by
Widrow~\cite{widrow89}, and analyze gravitational perturbations in the
interior solution.  We will display a master wave equation that
governs the evolution of the gravitational field modes.  This equation
is solved in the thin wall limit, yielding very interesting results
concerning the reflection and transmission of gravitational waves by
domain walls. 
This extends to gravitational waves  
the work presented in~\cite{vilenkinshellard} (see section
13.4 of the book) for the scattering of scalar field waves.
The thick wall is also considered and the properties of the
waves inside the wall are analyzed in what concerns their spin
content.

This type of analysis for the study of the interaction between matter
and gravitation has been performed in a previous work of ours for thin
systems with perfect fluid matter~\cite{bentolemos}.  Here, we apply
it to domain walls but including now the transition between matter and
vacuum.


\section{Background configuration}   \label{configuration}

The action for the scalar field is 
\begin{equation}
{\mathcal{S}}_{\varphi}=\int{ \sqrt{-g}\, {\mathcal{L}}_{\varphi}\, d^4 x}= 
\int{ \sqrt{-g} 
\left( \frac{1}{2}g^{\mu \nu }\partial_{\mu }\varphi
\, \partial _{\nu }\varphi -V(\varphi ) \right) d^4x }\;,  \label{lagrangian1}
\end{equation}
where $\mathcal{L}_{\varphi}$ is the scalar field Lagrangean, and $g$ 
is the determinant of the metric.
The action  for the gravitational field is 
\begin{equation}
{\mathcal{S}}_{g}= - \frac{1}{16\pi G}\int { \sqrt{-g}\,R\, d^4 x }\;,  
\label{lagrangian2}
\end{equation}
where $R$ is the Ricci scalar. 
The equations of motion, of the joint action $S_{\varphi}+S_{g}$,
for the scalar and gravitational fields are, respectively, 
\begin{equation}
\square \varphi +\frac{d V}{d \varphi }=0\;,
\label{kleingordon}
\end{equation}
\begin{equation}
R_{\mu \nu }=8\pi G(T_{\mu \nu }-\tfrac{1}{2}T_{\alpha }^{\alpha }\,g_{\mu
\nu })
\;,  						\label{einstein}
\end{equation}
where, 
$\square\equiv \frac{1}{\sqrt{-g}}\partial_\mu
\left[\sqrt{-g}g^{\mu\nu}\partial_\nu\,\right]$, 
$R_{\mu\nu}$ is the Ricci tensor, and $T_{\mu\nu}$ is the 
energy-momentum tensor given by
\begin{equation}
T_{\mu \nu }=\partial _{\mu }\varphi \,\partial _{\nu }\varphi -g_{\mu \nu }%
{\mathcal{L}}_{\varphi }
\; .  						\label{tmunu}
\end{equation}

The domain wall will be specified by the scalar and  gravitational 
field solutions $\bar{\varphi}$ and $\bar{g}_{\mu\nu}$, respectively. 
We assume that the domain wall has plane symmetry along the $x-y$ 
plane (i.e., the group of motions is $E_2$), 
and reflection symmetry under $z\rightarrow -z$. 
The absolute minimum value for the potential is assumed to be given by
$V_{\min }=0$.
The scalar field depends on $z$ only, $\varphi =\bar{\varphi}(z)$. 
$\bar{\varphi}(0)$ is a local maximum of the potential and 
$\varphi_+=\bar{\varphi}(+\infty )$, 
$\varphi_-=\bar{\varphi}(-\infty )$ are absolute minima of
the potential, $V(\bar{\varphi}(\pm \infty ))=V_{\min }=0$.
Reflection symmetry implies
$\bar{\varphi}(z)-\bar{\varphi}(0)=\bar{\varphi}(0)-\bar{\varphi}(-z)$. 
The potential is an even function of $z$ due to the reflection property, 
$V(\varphi )=V(2\bar{\varphi}(0)-\varphi )$, around the local maximum 
$\bar{\varphi}(0)$.
The gravitational field depends on 
time $t$ and coordinate $z$, $\bar{g}_{\mu \nu }(t,z)$. 
Reflection symmetry implies
$\bar{g}_{\mu \nu }(t,z)=\bar{g}_{\mu \nu }(t,-z)$.

The wall creates a gravitational field with a line element of the form 
\begin{equation}
d\bar{s}^{2}=D(z)\left( dt^{2}-dz^{2}-\expo^{-st}(dx^{2}+dy^{2})\right) \;,
\label{widrowmetric}
\end{equation}
in the coordinates used by Widrow~\cite{widrow89}, where $s\simeq 
4\pi G\sigma $ 
and $\sigma $ is the wall tension. 
The equations of motion (\ref{kleingordon}) and (\ref{einstein}) are written 
as 
\begin{subequations}
\label{background}
\begin{eqnarray}
\bar{\varphi}^{\prime \prime } +\frac{D^\prime}{D} \, \bar{\varphi}^{\prime} 
- D \, 
\frac{d V}{d \varphi }
(\bar{\varphi}) =0 
\;,  						\label{phibar} \\
({D^\prime}\!/{D} )^\prime +
 \frac{16\pi }{3}G \, (\bar{\varphi}^{\prime 2}+ D V(%
\bar{\varphi})) =0 
\;,  						\label{D}
\end{eqnarray}
\end{subequations}
where, $^\prime\equiv d\;/dz$. 
$D$ is an even function of $z$ and 
$D'/D$ varies from $0$ at $z=0$ to a constant value outside the wall
where $\bar{\varphi}', \; V  \rightarrow 0$.
More precisely~\cite{widrow89}, 
$D'/D \rightarrow \mp s$ as $z \rightarrow \pm \infty$.  
The function $D$ can and will be normalized to $D(0)=1$.

Turning off gravity, i.e., using $n_{\mu\nu}$ in Eq.\ (\ref{kleingordon}), 
the scalar field 
equation of motion admits a solution 
$\varphi _{0}(z)$ satisfying 
\begin{equation}
\varphi _{0}^{\prime 2}=2\, V(\varphi _{0})\;,\qquad \qquad \varphi
_{0}^{\prime \prime }=\frac{d V}{d \varphi }(\varphi _{0})\;.
							\label{phi0}
\end{equation}
The gravitational field produced by the wall back reacts producing 
the scalar field solution $\bar{\varphi}(z)$ instead of $\varphi _{0}(z)$.
The most often considered potentials are the $\varphi^4$ potential, 
\begin{equation}
V= \frac{1}{4}\, \lambda (\varphi^2 - \eta^2)^2
\;, 						\label{Vphi4}
\end{equation}
where $\lambda$ is a coupling constant and $\eta$ a mass scale, 
and the axionlike potential (see, e.g.,~\cite{vilenkinshellard})
\begin{equation}
V= \frac{m^2 \eta^2}{N^2}\, (1- \cos(N {\varphi}/{\eta}))
\; , 						\label{Vaxion}
\end{equation}
where $m$ is the mass of the pseudo-Nambu-Goldstone boson (axion), 
much smaller than the scale $\eta$, and $N$ is an integer.
In the first case the energy density of the wall is 
$\rho_0 \sim \lambda \eta^4$, the wall thickness is 
$\ell \sim \lambda^{-1/2} \eta^{-1}$ and the wall tension is 
$\sigma = \rho_0 \ell \sim \lambda^{1/2} \eta^3$.
In the second case, the same quantities are of the order of magnitude, 
respectively, 
$\rho_0 \sim m^2 \eta^2$, $\ell \sim m^{-1}$ and
$\sigma \sim m \, \eta^2$.
We are interested in perturbations with very large wavelengths up to the 
order of $\sim1/\sqrt{G\rho}$,
i.e., with wave numbers as small as $\kc$ defined as
\begin{equation}
\kcs = 4\pi G \rho_0
\;. 						\label{kcs}
\end{equation}
Notice the following hierarchy: $s/\kc \simeq \kc\, \ell \sim
\eta/M_{\rm P}$ is an extremely small number and $s\, \ell \simeq \kcs \,
\ell^2 \sim \eta^2/M^2_{\rm P}$ is even smaller.


\section{Gravitational perturbation equations}   \label{equations}

The background is thus given by $\bar{\varphi}(z)$
and $\bar{g}_{\mu \nu }(t,z)$. 
Consider now perturbations around this background configuration,
\begin{subequations} 
\label{phigwaves} 
\begin{eqnarray}
\varphi  &=&\bar{\varphi}(z)+\delta \varphi (x^\mu)\;,  
							\label{phi} \\
g_{\mu \nu } &=&\bar{g}_{\mu \nu }(t,z)+h_{\mu \nu }(x^\mu)\;.  	
								\label{gmunu}
\end{eqnarray}
\end{subequations} 
$\varphi $ and $g_{\mu \nu }$ obey the Klein-Gordon and Einstein equations 
(\ref{kleingordon}) and (\ref{einstein}).

We are going to assume that $\bar{g}_{\mu \nu }$ as given in 
(\ref{widrowmetric})
differs only slightly from the Minkowski metric $\eta _{\mu \nu}$ 
($\eta _{\mu \nu }=\mathrm{diag}(1,-1,-1,-1)$), i.e., 
$\bar{g}_{\mu \nu}-\eta _{\mu \nu }$ is a small quantity. 
This amounts to say that $s|t|\ll 1$, $s|z|\ll 1$ 
(which assures that $|D-1|\ll 1$ because $D(0)=1$ and $|D'/D| \leq s $).
In this region of the space-time the deviation of the scalar field solution 
$\bar{\varphi}(z)$ from $\varphi_0 (z)$ is suppressed by the Newton constant 
$G$.
To be consistent we treat the equations expanding up to a well defined  
order in $G$.

Subtracting from the equations obeyed by $\varphi(x^\mu)$ and 
$g_{\mu\nu}(x^\mu)$,
the equations obeyed by the background fields $\bar{\varphi}$ and 
$\bar{g}_{\mu \nu }$, one obtains the equations satisfied by
the perturbations $\delta \varphi $ and $h_{\mu \nu }$. 
In leading order in  $G$, 
the linearized equations in $\delta \varphi $ and $h_{\mu \nu }$ are, 
\begin{equation}
\eta ^{\mu \nu }\,\delta \varphi _{,\mu \nu }+\frac{d^{2}V}{d
\varphi ^{2}}(\varphi _{0})\,\delta \varphi =\frac{d V}{d
\varphi }(\varphi _{0})\,h_{zz}\;,  \label{dphi}
\end{equation}
\begin{equation}
R_{\mu \nu }-\bar{R}_{\mu \nu }=8\pi G\left( -\tfrac{1}{2}\varphi
_{0}^{{\prime }2}h_{\mu \nu }-\eta _{\mu \nu }\,\varphi _{0}^{\prime \prime
}\,\delta \varphi +\varphi _{0,\mu }\,\delta \varphi _{,\nu }+\varphi
_{0,\nu }\,\delta \varphi _{,\mu }\right) \;,  \label{dEinstein}
\end{equation}
where we have used  the de Donder (or harmonic)
gauge condition, 
\begin{equation}
\eta ^{\alpha \mu }h_{\mu \nu ,\alpha }-\frac{1}{2}\eta ^{\alpha \beta
}\,h_{\alpha \beta ,\nu }=0\;.  \label{Donder}
\end{equation}
In equations (\ref{dphi})-(\ref{dEinstein}) we have neglected 
terms that go as $(\bar{\varphi}-\varphi _{0})\delta \varphi 
$, $(\bar{\varphi}-\varphi _{0})h_{\mu \nu }$, $(\bar{g}_{\alpha \beta
}-\eta _{\alpha \beta })\delta \varphi $ or $(\bar{g}_{\alpha \beta }-\eta
_{\alpha \beta })h_{\mu \nu }$
 because they are of higher order in $G$. 
This is, of course, compatible with our assumption 
that the calculations are only valid within the range $|t|,\,|z|\ll
s^{-1}$, ($\,|D-1|\ll 1\,$). 
However, the nonlinearity of
Einstein equations is taken in account by keeping the terms in the first
member of Eq.\  (\ref{dEinstein}) that go as 
$(\bar{g}_{\alpha \beta }-\eta_{\alpha \beta })h_{\mu \nu }$,
which are comparable with 
$G \varphi_{0}^{{\prime }2}h_{\mu \nu }$ 
in the second member.

Equation (\ref{dphi}) is satisfied for modes with $\,h_{zz}=0$ and 
$\delta \varphi =0$. 
In this paper we study the gravitational modes obeying that condition, 
except in section \ref{modeplus} where a gauge transformation is performed
that modifies $\,h_{zz}=0$ and $\delta \varphi $.

Given the plane symmetry of the wall and the well
defined time scale, $s^{-1}$, of the background gravitational field, it is
legitimate to look for periodic wave modes of the form 
\begin{equation}
h_{\mu \nu }=A _{\mu \nu }(z)\,
\expo^{i (-\omega t + \kx x +  \ky y )}\;,
						\label{hmunu}
\end{equation}
as long as $\omega \gg s$.  In these conditions the explicit
dependence of Eq.\ (\ref{dEinstein}) on time and parameter $s$ can be
neglected, which we do.  Moreover, since the background curvature
$\sim \sqrt{G\rho }$ ($\rho =\sigma /\ell $ is the volume energy
density and $\ell $ the thickness of the wall) is much bigger than
$s\simeq 4\pi G\sigma $, $\sqrt{G\rho }\gg s$, one can still have
$\omega\sim \sqrt{G\rho }$. Thus, it is still possible to study
effects of gravitational modes on matter and vice versa.

Recently, perturbations of static domain walls in generic
n+1D space-times were studied in~\cite{kakushadze00} and it
was shown that the equations of motion admit solutions obeying the
additional constraints $h_{\mu z}=0$, $\eta^{\mu \nu} h_{\mu \nu}=0$.
In combination with the de Donder gauge condition (\ref{Donder}) they
define transverse traceless modes with respect to the coordinates $t,
x, y$, i.e.,
\begin{equation}
\eta^{a b} h_{b c, a} = 0 \; ,  \qquad \eta^{a b} h_{a b} = 0 \; 
 \qquad   (a, b= t,x, y)\; .
						\label{ttgauge}
\end{equation}
Without loss of generality we consider waves propagating along the 
$x-z$ plane, that is, $\ky = 0$ in Eq.\  (\ref{hmunu}).
There are two independent modes satisfying the above constraints,
that we name ``$\times$'' and  ``$+$'' modes for reasons that will be clear
in section \ref{spin}. 
The  ``$\times$'' mode obeys the following relation:
%
\begin{eqnarray}
\omega h_{t y} = -\kx h_{x y}  \;. 
				  	\label{hxy}
\end{eqnarray}
In turn, the  ``$+$'' mode obeys	
\begin{eqnarray}				
\omega^2 h_{t t} = -\omega \kx h_{t x}  = \kx^2 h_{x x}  \;, 	
								\label{hyy} 
\end{eqnarray}
which upon using the tracelessness condition $h_{yy}=h_{tt}-h_{xx}$
also yields, 
\begin{eqnarray}	
\omega^2 h_{y y} = (\kx^2 - \omega^2) h_{x x}  \;. 
\label{hyy2} 
\end{eqnarray}
Equation  (\ref{dEinstein}) gives for these ``$\times$'' and ``$+$'' modes,
\begin{equation}
(\omega ^{2}-\kx^{2}) h_{\mu \nu} + 
(D (D^{-1} h_{\mu \nu})^\prime )^\prime  =0 \;,
						\label{heqmotion}
\end{equation}
up to higher order terms in $G$.
In this equation it is taken into account that $D(z)$ relates with the energy 
density of the free wall, $\rho _{0}(z)$, as 
\begin{equation}
D^{\prime \prime }\simeq -8\pi G\rho _{0}=
 -8\pi G\varphi _{0}^{\prime 2}\;,
						\label{Dapprox}
\end{equation}
in accordance with eqs.\ (\ref{D}) and (\ref{phi0}).
Neglecting higher order terms in $G$ means here to consider 
$D^{\prime \prime} D^n \simeq D^{\prime \prime}$,
$D^\prime  D^n \simeq D^\prime $, $D^{\prime 2} \simeq 0$.
This will be consistently carried out throughout the paper.

We are interested in two main aspects, refraction and spin content of
the waves.  It is convenient, specially in the case $\omega ^{2}\neq
\kx^{2}$, to define new functions, 
\begin{equation}
\psi_{\mu \nu} = D^{-1/2} h_{\mu \nu} \;,
						  \label{psi} 
\end{equation}
in such a way that they obey a more familiar 
differential equation, 
\begin{equation}
(\omega ^{2}-\kx^{2})\psi_{\mu \nu} + {\psi^{\prime \prime}_{\mu \nu}} +
4 \pi G \rho_0\,\psi_{\mu \nu} =0\;.
						\label{psieqmotion}
\end{equation}
It has the form of a time-independent one-dimensional
Schr\"{o}dinger equation where $-4\pi G\rho _{0}$ takes the role of a
potential.  This equation can be solved analytically for particular
energy density profiles which will be done below.
Equation (\ref{psieqmotion})
is the master equation for the perturbations we have been
considering. Equivalently, one can use Eq.\ (\ref{heqmotion})
as our master equation.

In order to identify the spin of the waves it is necessary to perform
certain coordinate transformations.  Under the infinitesimal
transformation $x^\mu = \hat{x}^\mu + \xi^\mu$, the perturbations of the 
scalar and
gravitational fields given by Eqs.\ (\ref{phigwaves}) transform into
\begin{subequations}
\label{gaugetrf}
\begin{eqnarray}
\delta \hat{\varphi} \! & = & \! \delta \varphi + 
\bar{\varphi}^\prime {\xi}^z \;,
					\label{phi'} \\
\hat{h}_{\mu \nu }\! & = & \! {h}_{\mu \nu } + {\eta}_{\mu \nu } D^\prime 
\xi^z 
+ D ({\xi}_{\mu ,\nu} + {\xi}_{\nu ,\mu}) \;,
						\label{hmunu'}
\end{eqnarray}
\end{subequations}
where ${\xi}_{\mu}={\eta}_{\mu \nu } \xi^\nu$.
Here we neglect again the time dependence of the bakground metric 
(\ref{widrowmetric})
(recall that we have assumed $\omega \gg s$). The spin composition will 
be studied 
in detail in section \ref{spin}. 


\section{Solutions of the wave equation}  \label{solutions}

It is clear that the phase velocity in the wall is different from the
phase velocity in vacuum, $c=1$.  Consequently, there is a refraction
and reflection phenomenum when a wave enters the wall from outside or
emerges off the wall.  We analyze first the particular case $\omega
^{2}=\kx^{2}$, and then study the cases $\omega^{2}\neq \kx^{2}$.

\subsection{$\omega ^{2}=\kx^{2}$}  \label{solutions1}

 From Eq.\  (\ref{heqmotion}) one derives  two independent 
kinds of solutions,
\begin{eqnarray}
h_{\mu \nu} \!&=&\! \varepsilon_{\mu \nu}\, D(z)\, \expo^{-i\, 
\omega (t-x)} \;,
						\label{hsol1} \\
h_{\mu \nu} \!&=&\! \varepsilon_{\mu \nu}\, D(z) 
\int^z_0 \frac{d z'}{D(z')} \, \expo^{-i\, \omega (t-x)} \;,
						\label{hsol2}
\end{eqnarray}
where $\varepsilon_{\mu \nu}$ are constant polarization tensors constrained
either by Eq.\  (\ref{hxy}) or Eqs.\  (\ref{hyy}), (\ref{hyy2}).

The spin composition of the waves will be studied in section \ref{spin} in the general 
case. 
Neverthless, we study here briefly the spin content of
the particular waves in Eqs.\ (\ref{hsol1}) and (\ref{hsol2}).  We
resort to eqs.\ (\ref{gaugetrf}).  It turns out
that the solutions specified by Eqs.\ (\ref{hsol1}), (\ref{hxy}) and
(\ref{hyy})-(\ref{hyy2}) can be gauged away with appropriate transformations
$\xi^t$, $\xi^x$ and $\xi^y$.  That is not so for the solutions
(\ref{hsol2}).

A transformation
\begin{equation}
\xi^y = \frac{i}{\omega} D^{-1} h_{ty} 
						\label{epsilony}
\end{equation}
brings the wave specified by Eqs.\  (\ref{hxy}) and (\ref{hsol2}) into 
the canonic form where only the $\hat h_{yz}$ component is nonzero 
and its amplitude is constant everywhere:
\begin{equation}
\hat{h}_{yz} =- \frac{i}{\omega} \varepsilon_{ty}\, \expo^{-i\, \omega (t-x)} 
\; .						\label{hyz}
\end{equation}
It represents an helicity-2 gravitational wave that propagates parallel 
to the wall at the speed of light, inside and
outside the wall, without any noticeable interaction.
The other mode, identified by Eqs.\  (\ref{hyy})-(\ref{hyy2}), has the pathological 
feature of $h_{yy}=0$ in the limit $\omega^2=\kx^2$ and it is preferable 
to study the generic case $\omega^2 \neq\kx^2$ in section \ref{spin}.

\subsection{$\omega ^{2}\neq \kx^{2}$} \label{solutions2}

To obtain solutions for  eq.\ (\ref{psieqmotion}) one has to
present a given potential $\frac12 D^{\prime \prime } = - 4\pi G\rho
_{0}$.  Outside the wall, where the energy density is negligible,
Eq.\  (\ref{psieqmotion}) admits a plane wave solution,
\begin{equation}
\psi_{\mu \nu }=\varepsilon_{\mu \nu }\,
\expo^{i (-\omega t + \kx x +  \kz z )}\;,
						\label{psipw}
\end{equation}
propagating at the speed of light, i.e., with a dispersion relation 
given by 
\begin{equation}
\omega ^{2}=\kx^{2}+\kz^{2}  \, .
						\label{dispersion)}
\end{equation}
Inside the wall, if the density
$\rho_0$ is constant or varies sufficiently slowly, Eq.\
(\ref{psieqmotion}) also admits a plane wave solution in the interior
of the wall but with a different dispersion relation,
\begin{equation}
\omega ^{2}=\kx^{2}+\kz^{2} - 4\pi G\rho _{0} \, .
						\label{dispersion}
\end{equation}
A wave travelling into or from the wall gets different values of
momentum  along the $z$ direction inside and outside the wall, while
keeping the same value of $\kx$.  This is a typical refraction
phenomenum.  There are two aspects to analyze, one is the reflection and
transmission properties of the wall and the other 
the spin content.

The class of domain wall models we consider correspond to thin
walls in the sense that the product of the potential times the square
of the width $\ell$ is very small, i.e., $4 \pi G \rho_0 \ell^2
\simeq s \ell \sim \eta^2/M^2_{\rm P} \lll 1$.  
If, in addition, the wavelength ${k_z}^{-1}$ is much larger than the 
wall thickness, ${k_z}\ell\ll1 $,  it
is sufficient to consider the extreme thin wall limit to study the wave
spectrum and transition from the wall to the vacuum. 
That amounts to consider the
$\delta-$function wall, which gives rise to the Vilenkin wall, 
\begin{equation}
\rho _{0}=\sigma \,\delta (z)\,.
\label{vilenkinwall}
\end{equation}
Then, eq.\ (\ref{psieqmotion}) is now, omiting the tensor indices,
\begin{equation}
{\psi}'' + 4\pi G\sigma \, \delta (z)\,\psi + 
(\omega ^{2}-\kx^{2})\psi  =0\;.
\label{psieqmotion2}
\end{equation}
We stress that this equation applies to both polarization states
``$\times$'' and ``$+$''.
Its solutions can be further separated into two cases.

\begin{description}

\item[(a)] $\omega ^{2}-\kx^{2}<0$
\end{description}

This case corresponds to a `bound' state of the Schr\"{o}dinger equation. 
We write $\alpha ^2=-(\omega ^{2}-\kx^{2})$, with $\alpha^2 >0$. 
The solution of eq.\ 
(\ref{psieqmotion2}) is then
\begin{subequations}
\label{solutionlessthanzero} 
\begin{eqnarray}
\psi(z)=a\, \expo^{\alpha z} \;, \phantom{\expo^{-\alpha z}} \quad & z<0 \;, 
\label{negative} \\
\psi(z)=a\, \expo^{-\alpha z} \;, \phantom{\expo^{\alpha z}} \quad & z>0 \;. 
					\label{zpositive}
\end{eqnarray}
\end{subequations}
Integrating through the wall, one obtains $\alpha =2\pi G\sigma \simeq
s/2$, which yields the dispersion relation  $\omega^2 = \kx^2 - s^2
/4$.  However, since we have been assuming that $\omega\gg s$ and have
neglected $s$ in the linearized perturbation equations, this solution
cannot be differentiated from the cases $\omega ^{2}=\kx^{2}$ analyzed
before representing waves propagating along the plane of the wall.

 From the dispersion relation (\ref{dispersion}) one can draw general
conclusions, even when one does not restrict to a particular
potential, such as the one in Eq.\ (\ref{vilenkinwall}). Indeed, the
dispersion relation (\ref{dispersion}) suggests that that are
solutions where the amplitude does not vary in the interior to the
wall ($\kz=0$) corresponding to negative values of $\omega^2 -\kx^2 =
- 4\pi G\rho_0 $ in particular $\omega^2 < 0 $.  However, those wave
functions diverge exponentially in the space exterior to the wall with
a scale $\sqrt{4\pi G\rho_0} \gg s$ and do not provide well behaved
solutions.  The analogous problem for a fluid wall was treated in 
\cite{bentolemos}.
In view of the similarity between the dispersion relation
(\ref{dispersion}) and the dispersion relation obtained in the case of
a fluid wall for wave functions independent of $z$ (see also the mode given
by Eq.\ (\ref{hxy}) in this paper and the equivalent mode in~\cite{bentolemos}), 
it is likely that the same conclusion applies in
the case of the fluid.  In other words, the unstable modes, $\omega^2
< 0 $, diverge exponentially with the distance to the wall and
therefore do not belong to the actual wave spectrum when the full
spatial variation is taken into account.

\begin{description}

\item[(b)] $\omega ^{2}-\kx^{2}>0$

\end{description}

This case is a typical scattering problem. 
We write  $\kz ^2=\omega ^{2}-\kx^{2}$, with $\kz , \kz^2 >0$. The solution 
of eq.\ (\ref{psieqmotion2}) is
\begin{subequations}
\label{solutiongreaterthanzero}
\begin{eqnarray}
 \psi(z)=& a\, \expo^{i\kz z} + b\, \expo^{-i\kz z}   \;, 
 \qquad  & z<0 \;, 
				 \label{incident} \\
 \psi(z)=& c\, \expo^{i\kz z} \phantom{+ b\, \expo^{-i\kz z}} \;, \qquad  
& z>0 \;. 
				\label{transmitted}
\end{eqnarray}
\end{subequations}
Continuity and integration through the wall yield for the reflected and 
transmitted wave amplitudes, respectively,
\begin{equation}
b=-\frac{s}{s+2i\kz }\, a \;, \qquad c=\frac{2i\kz }{s+2i\kz }\, a\;.
\label{constants}
\end{equation}
Thus, the reflection ${\rm Re}$ and transmission ${\rm Tr}$ coefficients are
\begin{subequations}
\label{coefficients}
\begin{eqnarray}
{\rm Re} \equiv 
\frac{|b|^2}{|a|^2}=\frac{s^2}{s^2+4\kz ^2}\;,
						\label{re} \\
{\rm Tr} \equiv 
\frac{|c|^2}{|a|^2}=\frac{4\kz ^2}{s^2+4\kz ^2}\;. 
						\label{tr}
\end{eqnarray}
\end{subequations}
The ratio of both coefficients is ${\rm Re}/{\rm Tr}= {s^2}/{4\kz
^2}$.  Thus, for $\kz $ small (large wavelengths) compared to
$s\simeq4\pi G\sigma$, almost all of the wave is reflected. For $\kz $
large almost all of the wave passes through the wall. This conclusion
is similar to the one displayed in paragraph 13.4 of
\cite{vilenkinshellard} (see also~\cite{everett}) for the scattering of
scalar particles by domain walls without the inclusion of gravitational
effects. 
Notice that the reflection and transmission coefficients are the same for 
both the polarization states ``$\times$'' and ``$+$''.

In Eqs.\ (\ref{solutiongreaterthanzero})
we have dropped the indices of the wave function $\psi_{\mu\nu}$ and 
consequently of  $a_{\mu\nu}$, $b_{\mu\nu}$ and $c_{\mu\nu}$. 
However, the indice content is important when one wants to analyze the spin 
and helicity of the waves. 
Since, one can do this directly from the equations of motion of the waves 
(\ref{hxy}) and (\ref{hyy})-(\ref{hyy2}) in their most general form,
Eqs.\ (\ref{heqmotion}), 
there is no need to discuss for this particular $\delta-$function potential, 
one can study the spin conposition generically. 
We do this in the next section.


\section{Spin composition} \label{spin}

Gravitational waves only have helicity $\pm 2$ states in vacuum.  One
usually gauges away the time components of the wave tensor $h_{\mu
\nu}$, that correspond to spin 0 and 1 states, and verifies that the
remaining spatial tensor $h_{i j}$ is traceless and transverse.  That
eliminates the spin 0 and spin 2 states with helicity different from
$\pm 2$.  We do the same here for the modes identified in
Eqs.\ (\ref{hxy}) and (\ref{hyy})-(\ref{hyy2}) by applying sucessive gauge
transformations as expressed in Eqs.\ (\ref{gaugetrf}).

One point to keep in mind is that $D'$ varies from $0$ at $z=0$ to
$|D'| \simeq  4\pi G \rho_0 \ell \simeq  s$ on the surfaces of the
wall, $|z| \approx \ell /2$, as results from the integration of
Eq.\ (\ref{D}).  In addition, recall that $s \ll \sqrt{G \rho_0}$ and
$s \ll \omega$ hence, we will neglect $D'$ with respect to $\omega$ and
$\sqrt{G \rho_0}$ but not $D'' \simeq  -8\pi G \rho_0$ with respect to
$\omegas$.  

Our first order expansion in the Newton constant forces us
to neglect products such as $G \rho_0 D'$ or $(G \rho_0)^2$ in section 
\ref{modeplus}.
For the mode ``$+$'' studied in that section it is
convenient to consider a simple energy density profile that is almost
flat around its maximum which is true for some scalar field potential
models.  
Alternatively one may consider that the calculations
apply only to the points where the energy density has a maximum, or
that $D''' =0$, which indeed occurs in the central plane of the wall,
$z=0$.

\subsection{Mode ``$\times$''}     \label{modetimes}

This mode obeys the relation (\ref{hxy}) and the equations of motion 
(\ref{heqmotion}).
Applying the transformation $\xi^y$ given in Eq.\ (\ref{epsilony}),
the component $h_{ty}$ is gauged away and one obtains 
\begin{subequations}
\label{barhtimes}
\begin{eqnarray}
\hat{h}_{xy} &=& \frac{\omegas -\kxs}{\kxs}\, h_{xy} 
\;,						\label{barhxy} \\
\hat{h}_{yz} &=&  \frac{i\, \kx}{\omegas}\, D (D^{-1} h_{xy})'
\;. 						\label{barhyz}
\end{eqnarray}
\end{subequations}
 From this and the $h_{xy}$ equation of motion (\ref{heqmotion}) one 
derives the relation,
\begin{equation}
i \, \kx \, \hat{h}_{xy} + \hat{h}'_{yz} = 0
\; , 						\label{Ttimes}
\end{equation}
and also the equation of motion of $\hat{h}_{yz}$,
\begin{equation}
(\omega ^{2}-\kx^{2}) \hat{h}_{yz} + 
D (D^{-1} \hat{h}'_{yz})'  =0
\; . 						\label{barhyzeqmotion}
\end{equation}
Note that this is different from the equation of motion of ${h}_{xy}$ and 
$\hat{h}_{xy}$.

This mode is obviously traceless and is also transverse because, as 
results from Eq.\ (\ref{Ttimes}),
\begin{equation}
\bar{\nabla}^{\mu} \hat{h}_{\mu \nu} \simeq
D^{-1} \eta^{\mu \alpha} \partial_\alpha \hat{h}_{\mu \nu} =0
\; , 						\label{cTtimes}
\end{equation}
where the covariant derivative refers to the unperturbed metric
$\bar{g}_{\mu \nu}$.  The terms neglected in the first equality are
suppressed by $|D'| /\omega \lesssim s /\omega \ll 1$.  The conclusion
is that this particular gravitational wave constitutes a well defined
polarization state ``$\times$'' of helicity $\pm 2$ even in the interior
of the domain wall where the refraction index is modified (see
Eqs.\ (\ref{psieqmotion}) and (\ref{dispersion})).

\subsection{Mode ``$+$''}  \label{modeplus}

This mode is specified by Eqs.\ (\ref{hyy})-(\ref{hyy2}). 
The components ${h}_{tt}$ and ${h}_{tx}$ are gauged away with the 
transformations
\begin{subequations}
\label{epsilontx}
\begin{eqnarray}
\tilde\xi^t &=& - \frac{i}{2\omega}\, D^{-1} h_{tt}
\;,						\label{epsilont} \\
\tilde\xi^x &=& \frac{i\, \kx}{2 \omegas} \,
\frac{2\omegas - \kxs}{\omegas -\kxs} D^{-1}\, h_{yy}
\;. 						\label{epsilonx}
\end{eqnarray}
\end{subequations}
The new tensor is
\begin{subequations}
\label{tildecross}
\begin{eqnarray}
 \tilde{h}_{tz} &=& \frac{i}{2\omega}\, \frac{\kxs}{\omegas -\kxs}\,
D (D^{-1} h_{yy} )'
\;,						\label{tildehtz} \\
 \tilde{h}_{xx} &=& - \frac{\omegas-\kxs}{\omegas}\, h_{yy}
\;, 						\label{tildehxx} \\
 \tilde{h}_{yy} &=& {h}_{yy}
\;, 						\label{tildehyy} \\
 \tilde{h}_{xz} &=& - \frac{2\omegas - \kxs}{\omega \kx}\,
\tilde{h}_{tz}
\;. 						\label{tildehxz}
\end{eqnarray}
\end{subequations}
Using Eq.\ (\ref{heqmotion}) one gets
\begin{equation}
 \tilde{h}'_{tz} = -\frac{i\, \kxs}{2 \omega} h_{yy}
\;. 						\label{tildehtzhyy}
\end{equation}
$\tilde{h}_{tz}$ obeys the same wave equation  (\ref{barhyzeqmotion}) 
as $\hat{h}_{yz}$. 
To eliminate $\tilde{h}_{tz}$ one needs an $\xi^z$ transformation
which introduces a perturbation in the scalar field as follows from
Eq.\ (\ref{phi'}).  There is no other way of gauging away all the time
components $h_{t\mu}$ without introducing $\xi^z$.  This already shows
that this mode has a scalar, spin 0 component and is not a pure spin 2
wave.  To keep $h_{t t}$ and $h_{t x}$ equal to zero one has to
perform a simultaneous transformation as follows:
\begin{subequations}
\label{epsilontxz}
\begin{eqnarray}
 \xi^t &=&  - \frac{i}{2\omega}\, D'\, D^{-1} \xi^z  \;,	
\phantom{-\frac{ \kx}{ \omega} \, \xi^t 99999999}		
		\label{epsilontz} \\
 \xi^x &=& -\frac{ \kx}{ \omega} \, \xi^t  \;,
\phantom{\frac{i}{2\omega}\, D'\, \xi^z 99999999}
 						\label{epsilonxz} \\
\xi^z &=&  \frac{1}{\omegas + \kcs} \, D^{-1} 
\left( i\, \omega \,\tilde{h}_{tz} + \frac{1}{4}\,
\frac{\kxs}{\omegas + 2 \kcs}\, D' h_{yy} \right)
\; , 						\label{epsilonz}
\end{eqnarray}
\end{subequations}
where $\kcs$ is defined in Eq.\ (\ref{kcs}).  Here it is important to
assume that $D'''$ and $\rho'_0$ are negligible small which is true in
the central plane of the wall where they vanish, or in any model where
the scalar field potential is flat in a certain interval around its
maximum.  Making use of Eqs.\ (\ref{tildehtz}) and (\ref{tildehtzhyy})
the new wave tensor comes as ($\hat{h}_{tz}=0$)
\begin{subequations}
\label{barcross}
\begin{eqnarray}
 \hat{h}_{xx} &=& -\frac{\omegas -\kxs}{\omegas} 
\left(h_{yy} + \frac{i\, \omega}{\omegas +\kcs}\,
D'\, \tilde{h}_{tz} \right)
\;,						\label{barhxx} \\
 \hat{h}_{yy} &=& {h}_{yy}
\;, 						\label{barhyy} \\
 \hat{h}_{zz} &=& \frac{1}{\omegas +2\kcs} \left(-\kxs \, {h}_{yy} +
 i\, \omega \frac{2 \omegas -\kxs }{\omegas +\kcs} 
 D'\, \tilde{h}_{tz}  \right)
\; , 						\label{barhzz} \\
\hat{h}_{xz} &=& \frac{-2}{\omegas +\kcs} 
\left( \frac{\omega}{\kx}   (\omegas -\kxs +\kcs) \tilde{h}_{tz} 
+ \frac{i}{4}\, \frac{\kx^3}{\omegas
+2\kcs}\, D'\, h_{yy} \right) 
\;, 							\label{barhxz} 
\end{eqnarray}
\end{subequations}
with $\tilde{h}_{tz}$ given by Eq.\ (\ref{tildehtz}).
There is a spin 0 scalar field perturbation
\begin{equation}
\delta \hat{\varphi} = \bar{\varphi}^\prime \xi^z
\;. 						\label{bardphi}
\end{equation}

There is also a gravitational field spin 0 term, identified with the 
trace of $\hat{h}_{\mu\nu}$. Indeed, the trace is given by
\begin{equation}
\hat{h} = \hat{h}_{xx} + \hat{h}_{yy} + \hat{h}_{zz} =
\frac{2 \kcs}{\omegas + 2 \kcs}\, \frac{\kxs}{\omegas}\, \hat{h}_{yy} + \cdots 
\;. 						\label{trace}
\end{equation}
The ellipses represent terms suppressed by $|D'| /\omega \lesssim s/\omega
\ll 1$.  Remarkably enough the trace is proportional to $\kcs = 4\pi G
\rho_0$ showing that it only exists in matter, in the domain wall,
vanishing in the surrounding empty space.  

There are no spin 1 waves, since $\hat{h}_{ij}$ is symmetric. 
To determine if there are
spin 2 states with helicity different from $\pm 2$ one evaluates the
divergence of the traceless part
\begin{subequations}
\label{hT}
\begin{eqnarray}
\hat{h}^T_{ij} &=& \hat{h}_{ij} - \frac{1}{3}\, \hat{h}\, \delta_{ij}
\;, \qquad  i,j \neq t 
\; , 						\label{hTij}  \\
\hat{h}^T_{t \mu} &=& 0
\;. 						\label{hT0mu}
\end{eqnarray}
\end{subequations}
The result\footnote{If we had not neglected $D'''$ one would get
terms with the fourth derivative $D^{(4)}$, but that would introduce
one more model dependent parameter and could not change the
conclusions.} is again proportional to $\kcs$:
\begin{subequations}
\begin{eqnarray}
\label{cTcross}
\bar{\nabla}^{\mu} \hat{h}^T_{\mu t} &=& 0
\;, 						\label{cTcrosst} \\
\bar{\nabla}^{\mu} \hat{h}^T_{\mu x} & \simeq &
-\frac{4}{3}\, i\,  \kx \frac{ \kcs}{\omegas + 2 \kcs}\, 
\frac{\kxs}{\omegas}\, D^{-1}  \hat{h}_{yy} \;, 						\label{cTcrossx} \\
\bar{\nabla}^{\mu} \hat{h}^T_{\mu y} &=& 0
\;, 						\label{cTcrossy} \\
\bar{\nabla}^{\mu} \hat{h}^T_{\mu z} &\simeq &
-\frac{1}{3}\,  i\, \kx \kcs 
\frac{9\omegas -4\kxs}{(\omegas +2\kcs)(\omegas -\kxs+\kcs)}
\, D^{-1} \hat{h}_{xz}
\;. 						\label{cTcrossz}
\end{eqnarray}
\end{subequations}
It remains to tell whether the helicity is 0 or $\pm 1$. 
A spin 2 helicity 0 state corresponds to a divergence vector
$\bar{\nabla}^{\mu} \hat{h}^T_{\mu i}$ parallel to the wavevector and 
the helicity 1 
to a divergence vector orthogonal to the wavevector.
Here the wavevector has to be replaced by the more general 
quantity, the gradient $\bar{\nabla}^{\nu} 
\hat{h}^T_{\alpha \beta}$. Then, 
the signature of helicity 0 may be constructed as the internal product
\begin{equation}
H^0_{\alpha \beta}  = \bar{\nabla}^{\mu} \hat{h}^T_{\mu \nu} \, 
\bar{\nabla}^{\nu} 
\hat{h}^T_{\alpha \beta}
\;, 						\label{helicty0}
\end{equation}
and the helicity 1 signature as
\begin{equation}
H^1_{i j \alpha \beta }  = \bar{\nabla}^{\gamma} \hat{h}^T_{\gamma i} \, 
\bar{\nabla}_{j} \hat{h}^T_{\alpha \beta} -
\bar{\nabla}^{\gamma} \hat{h}^T_{\gamma j}\, \bar{\nabla}_{i} 
\hat{h}^T_{\alpha \beta}
\;. 						\label{helicty1}
\end{equation}
As expected none of the tensors vanish, for instance, 
$H^0_{yy} \simeq -3 \kcs \, \kxs \, \omega^{-2} \hat{h}^2_{yy}$ and 
$H^1_{xzxz} \simeq -5/3\, \kcs \, \kx^4 \, \omega^{-4} \hat{h}^2_{yy}$ 
are different from zero to first order in $G$.  
This completes the proof that a
perturbation that is a pure gravitational wave off the wall (spin 2
and helicity $\pm 2$), couples in the interior of the domain wall with
spin 0 and spin 2 degrees of freedom with helicities 0, $\pm 1$, 
and $\pm 2$.
Notice that although the results derived after Eqs.\ (\ref{epsilontxz}) 
are only strictly valid in the core of the wall, 
the mere existence of helicity and spin states other than
2 is a continuous property and therefore extends to the entire wall.


\section{Conclusions} \label{conclusions}

We have considered thick domain walls, mainly those studied by Widrow
\cite{widrow89}, and have analyzed gravitational perturbations in the
interior solution. We have displayed a perturbation equation which
yields the evolution of gravitational waves.  
We were able to solve
this equation in the thin wall limit, yielding
 results concerning the reflection and transmission properties of the
gravitational waves themselves that are similar to the ones obtained
for the scattering of scalar particles off domain walls presented in
\cite{vilenkinshellard}, where gravitational effects were not taken
into account.
The reflection and transmission coefficients are independent from the
gravitational wave polarization state.

The analysis of the spin content of the waves in the interior of the
wall shows that the two perturbation modes studied have different
properties.  One corresponds to a pure state of helicity $\pm 2$,
usually labeled with the symbol ``$\times$''.  The other mode reduces in
vacuum to the orthogonal state  with helicity $\pm 2$, denoted as
``$+$'', but in the interior of the wall it also contains a scalar field
perturbation and gravitational components with spin 0 and spin 2 states
of helicity $\pm 1$ and 0 whose  amplitudes are of the order of $4\pi G
\rho_0 /\omega^2$.

\section*{Acknowledgments}

We thank Funda\c c\~ao para a  Ci\^encia e Tecnologia (FCT) for  grants No.\
\newline 
CERN/P/FIS/40129/2000 and PESO/PRO/2000/40143.  L. B. thanks the
hospitality of the CERN Theory Division where part of this work was
done. J. P. S. L. thanks Observat\'orio Nacional do Rio de Janeiro for
hospitality.


\begin{thebibliography}{99}


\bibitem{vilenkinshellard} A. Vilenkin, E. P. S. Shellard, 
                  {\it Cosmic strings and other topological defects} 
(Cambridge University Press, Cambridge, England, 1994).
\bibitem{kibble} T. W. B. Kibble, J. Phys. A {\bf 9}, 1387 (1976).  
\bibitem{vilenkin1} A. Vilenkin, Phys. Rev. D {\bf 23}, 852 (1981).
\bibitem{vilenkin2} A. Vilenkin, Phys. Lett. {\bf 133B}, 177 (1983). 
\bibitem{ipsersikivie} J. Ipser, P. Sikivie, Phys. Rev. D {\bf 30}, 712 (1984).
\bibitem{hillschrammfry} C. T. Hill, D. N. Schramm, J. N. Fry,
                          Comm. Nucl. Part. Sci. {\bf 19}, 25 (1989). 
\bibitem{vilenkin3} A. Vilenkin, Phys. Rev. Lett. {\bf 72}, 3137 (1994).
\bibitem{rubakov83} V. A. Rubakov, M. E. Shaposhnikov, 
		Phys. Lett. {\bf 125B}, 136 (1983).
\bibitem{randallsundrum} L. Randall, R. Sundrum, Phys. Rev. Lett. 
                      {\bf 83}, 3370 (1999). 
\bibitem{akama00} K. Akama,  in {\em Proceedings of the Symposium on
Gauge Theory and Gravitation}, Nara, Japan, 1982, 
edited by K. Kikkawa, N. Nakanishi and
H. Nariai (Springer-Verlag, Berlin, 1983) (hep-th/0001113).
\bibitem{everett} A. E. Everett, Phys. Rev. D {\bf 10}, 3161 (1974). 
\bibitem{bonjour1} F. Bonjour, C. Charmousis, R. Gregory, Phys. Rev. 
                      D {\bf 62}, 083504 (2000). 
\bibitem{widrow89} L. M. Widrow, Phys. Rev. D {\bf 39}, 3571 (1989).
\bibitem{goetz}  G. Goetz, J. Math. Phys. {\bf 31}, 2683 (1990).
\bibitem{ishibashi} Y. Morisawa, R. Yamazaki, D. Ida, A. Ishibashi, K. Nakao,
                    Phys. Rev. D {\bf 62}, 084022 (2000). 
\bibitem{bonjour2} F. Bonjour, C. Charmousis, R. Gregory,
                      Class. Quant. Grav. {\bf 16}, 2427 (1999). 
\bibitem{bentolemos} L. Bento, J. P. S. Lemos,
                    Class. Quant. Grav. {\bf 17}, L117 (2000).
\bibitem{kakushadze00} Z. Kakushadze, P. Langfelder, Mod. Phys. Lett. A 
{\bf 15}, 2265 (2000).


\end{thebibliography}
\end{document}